\documentclass[twocolumn]{aastex62}
\usepackage{amssymb}
\usepackage{array,multirow}
\usepackage{comment}
\usepackage{enumerate}
\usepackage{bm}

\newcommand{\gaia}{{\it Gaia}}
\newcommand{\hipparcos}{{\it HIPPARCOS}}

\newcommand{\porb}{{P_{\rm{orb}}}}

\newcommand{\Porb}{\ifmmode {P_{\rm orb}}\else${P_{\rm orb}}$\fi}

\newcommand{\Msun}{\ifmmode {{M_\odot}}\else{$M_\odot$}\fi}
\newcommand{\Mtot}{\ifmmode {{M_{\rm tot}}}\else{$M_{\rm tot}$}\fi}
\newcommand{\RV}{\ifmmode {{\rm RV}}\else RV \fi}
\newcommand{\bigG}{\ifmmode {\mathcal{G}}\else${\mathcal{G}}$\fi}

\submitjournal{ApJ}

\shorttitle{Weighing the Darkness}

\begin{document}

\title{{\bf Weighing the Darkness II: Astrometric Measurement of Partial Orbits with \gaia}}

\author[0000-0001-5261-3923]{Jeff J. Andrews}
\affiliation{Center for Interdisciplinary Exploration and Research in Astrophysics (CIERA), 
1800 Sherman Ave., 
Evanston, IL, 60201, USA}
\affiliation{Niels Bohr Institute, University of Copenhagen, Blegdamsvej 17, 2100 Copenhagen, Denmark}
\email{jeffrey.andrews@northwestern.edu}

\author[0000-0002-9660-9085]{Katelyn Breivik}
\affiliation{Center for Computational Astrophysics, Flatiron Institute, 162 Fifth Ave, New York, NY, 10010, USA}
\affiliation{Canadian Institute for Theoretical Astrophysics, University
of Toronto, 60 St. George Street, Toronto, Ontario, M5S 1A7,
Canada}

\author[0000-0001-9685-3777]{Chirag Chawla}
\affiliation{Tata Institute of Fundamental Research, Department of Astronomy and Astrophysics, Homi Bhabha Road, Navy Nagar, Colaba, Mumbai, 400005, India}

\author[0000-0003-4175-8881]{Carl L. Rodriguez}
\affiliation{Harvard Institute for Theory and Computation, 
60 Garden St, 
Cambridge, MA 02138, USA}
\affiliation{McWilliams Center for Cosmology and Department of Physics, Carnegie Mellon University, Pittsburgh, PA 15213, USA}

\author[0000-0002-3680-2684]{Sourav Chatterjee}
\affiliation{Tata Institute of Fundamental Research, Department of Astronomy and Astrophysics, Homi Bhabha Road, Navy Nagar, Colaba, Mumbai, 400005, India}

\begin{abstract}
Over the course of several years, stars trace helical trajectories as they traverse across the sky due to the combined effects of proper motion and parallax. It is well known that the gravitational pull of an unseen companion can cause deviations to these tracks. Several studies have pointed out that the astrometric mission \gaia\ will be able to identify a slew of new exoplanets, stellar binaries, and compact object companions with orbital periods as short as tens of days to as long as \gaia's lifetime. Here, we use mock astrometric observations to demonstrate that \gaia\ can identify and characterize black hole companions to luminous stars with orbital periods longer than \gaia's lifetime. Such astrometric binaries have orbital periods too long to exhibit complete orbits, and instead are identified through curvature in their characteristic helical paths. By simultaneously measuring the radius of this curvature and the orbital velocity, constraints can be placed on the underlying orbit. We quantify the precision with which \gaia\ can measure orbital accelerations and apply that to model predictions for the population of black holes orbiting stars in the stellar neighborhood. Although orbital degeneracies imply that many of the accelerations induced by hidden black holes could also be explained by faint low-mass stars, we discuss how the nature of certain putative black hole companions can be confirmed with high confidence using \gaia\ data alone.
\end{abstract}


\section{Introduction}
\label{S:intro}

Over the course of several years, stars trace helical trajectories as they traverse across the sky due to the combined effects of proper motion and parallax. The satellite mission \hipparcos\ demonstrated the power of a dedicated astrometric instrument \citep{perryman1997, Hipparcos_catalog, van_leeuwen2007}, a paragon that was followed two decades later by its successor, \gaia\ \citep{Gaia_mission}. The second \gaia\ data release contains 1.7 billion stars, of which 1.3 billion have measured parallaxes and proper motions \citep{Gaia_DR2_summary, Gaia_astrometry}. The unprecedented scale and precision of these kinematic data are already revolutionizing our understanding of Galactic structure \citep[e.g.,][]{hunt2018}, stellar clusters \citep[e.g.,][]{cantat-gaudin2018}, and stellar binaries \citep[e.g.,][]{Kervella2019a}.

Since approximately half of all stars are found in stellar binary systems \citep{raghavan2010}, \gaia's impact on binary astrophysics is particularly consequential. For systems with orbital periods of minutes to days, these are astrometrically unresolved by \gaia, detectable by several methods \citep[e.g., photometric variability;][]{Gaia_variability}. For somewhat longer orbital periods, ranging from weeks to a few years, \gaia\ can astrometrically resolve these orbits, as their orbital motion can be identified as a residual on top of the parallactic and proper motion of a star \citep{Kervella2019a, andrews2019b, belokurov20, penoyre20}. At the opposite end, orbits with the longest periods can be identified as separate luminous stars moving together \citep{oelkers2017, oh2017, andrews2017, el-badry2018a}. Such common proper motion pairs can be used to constrain the initial-final mass relation \citep{andrews2015}, chemical tagging \citep{andrews2018, andrews2019a}, and the nature of gravity in the weak acceleration regime \citep{pittordis2018}. 

Between these regimes, a middle ground exists, where a binary companion can affect the trajectory of a star, but with an orbital period longer than \gaia's lifetime. These systems can be identified by having an accelerating motion on the sky, so that a star does not match the linear proper motion expectation for single, free-floating stars in the Galaxy \citep{Kervella2019a, belokurov20,penoyre20}. The detection of these astrometric binaries was addressed in the context of \hipparcos\ in a series of papers \citep{martin97, martin98a, martin98b}. \hipparcos\ produced a small catalog of 2622 stars that showed time-dependent proper motions \citep{falin99}, many of which were later identified as stellar binaries using follow-up spectroscopy and high-angular-resolution imaging \citep[e.g.,][]{mason01}.

The current publicly available \gaia\ data releases do not include the accelerations for stars exhibiting non-linear proper motions. However, recently \citet{Brandt2018} compared proper motions as measured by both \hipparcos\ and \gaia, producing a catalog of stars that exhibit astrometric accelerations. These have been used to constrain the orbits of a number of known and suspected binary stars and brown dwarf and planet hosts \citep{snellen18, Brandt2018b, torres19}. Although this data set includes only the subset of \gaia\ stars that were observed with both \hipparcos\ and \gaia, it exemplifies the value that such data can bring once it becomes publicly available for the full catalog of $>$10$^9$ stars observed by \gaia.

While there is clearly much to be learned about astrometric binaries in which both stars are luminous, here we are interested in the rarer cases of luminous stars hosting dark companions such as black holes (BH). Several previous studies have focused on the population of unseen BH binaries with orbital periods shorter than \gaia's lifetime \citep{Mashian2017, Breivik2017, Yamaguchi2018, Yalinewich2018, yi19, shahaf2019, wiktorowicz20, shikauchi2020}. These studies have been motivated in part by recent claimed detections of such detached BH binaries, both in the field \citep{thompson2019, Liu2019, rivinius2020,Jayasinghe2021} and in globular clusters \citep{Giesers2018, Giesers2019}. Though we note that, in practice, confirming BH binaries in the absence of astrometric information alone has proven difficult \citep[e.g., ][]{Abdul-Masih2020,El-Badry2020,Eldridge2020}.

In \citet[][hereafter Paper I]{andrews2019b}, we use mock \gaia\ observations to calculate the precision with which \gaia\ can measure the mass of dark companions to luminous stars in astrometric orbits. We find that the precision quickly degrades for orbits with periods longer than \gaia's lifetime, since only part of the orbit is observed. However, recent detailed binary evolution simulations \citep{langer20} as well as binary population synthesis predictions \citep{Breivik2017} demonstrate that a subset of luminous stars ought to have BH companions at orbital periods $\gtrsim$10 years. Even if we cannot completely solve for the orbits of these systems, we wish to address how well \gaia\ can measure the accelerations in these systems and whether those containing dark companions can be uniquely identified. As yet, we are unaware of any work that focuses on the population of BH binaries with orbital periods longer than \gaia's lifetime. 

In Section \ref{sec:detectability} we describe how \gaia\ detects such long period binaries, and we quantify its ability to measure orbital acceleration using mock \gaia\ observations. In Section \ref{sec:theory} we describe the orbital acceleration exhibited by stellar binaries in both circular and eccentric orbits. We apply those results to population synthesis predictions of BH binaries in Section \ref{sec:pop_synth}. Finally, we discuss our results and conclude in Section \ref{sec:conclusions}.

\section{Gaia's Detectability}
\label{sec:detectability}

\begin{figure}
    \begin{center}
    \includegraphics[width=1.0\columnwidth]{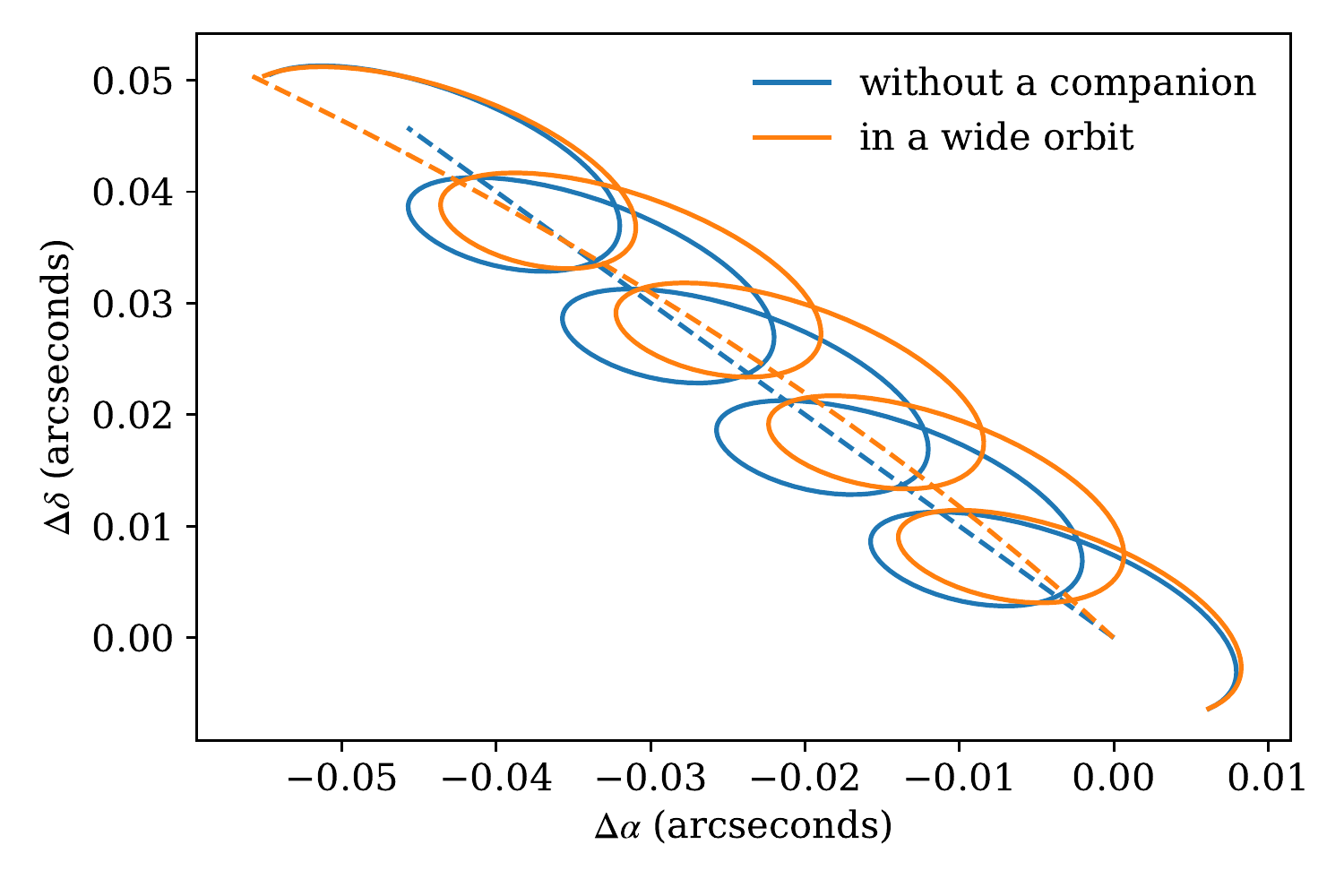}
    \caption{ Whereas an isolated star will form a helix as it moves across the sky (from both orbital parallax and proper motion), an unseen companion will induce an arc in this helical trajectory. }  
    \label{fig:orbital_motion}
    \end{center}
\end{figure}

The positions of isolated stars are subject to two effects: parallax and proper motion. While proper motion causes stars to move across the sky over time, parallax induces a circular offset depending on the time of year and position of the star in the sky. The resulting effect is that all stars produce helical trajectories as they move across the sky over time (see solid blue line in Figure \ref{fig:orbital_motion}). 

\begin{figure}
    \begin{center}
    \includegraphics[width=1.0\columnwidth]{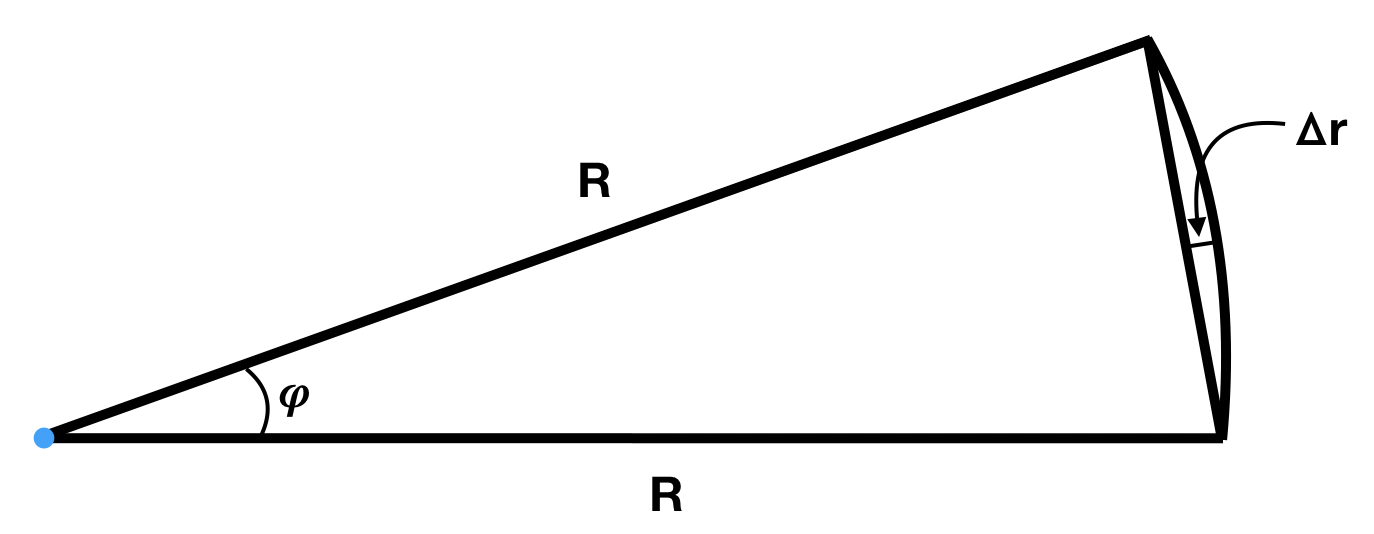}
    \caption{ \gaia's ability to detect orbital acceleration depends on measuring $\Delta r$, the radial difference between a chord and an arc. }  
    \label{fig:detectability_diagram}
    \end{center}
\end{figure}

If, however, a star feels the gravitational pull from an unseen object, the normally straight-lined helix will arc (see solid, orange line in Figure \ref{fig:orbital_motion}), with a magnitude depending on the acceleration it feels. Therefore \gaia's ability to measure the magnitude of an arc depends on its ability to discern between the arc of a circle and its corresponding chord\footnote{The detectability is derived for face-on orbits. Although it may not be immediately obvious, we demonstrate later in this Section that perspective effects from the orbit's orientation in space are in general quite small.}. For a circle of radius $R$, to lowest order the difference between these two lengths at any given observation time, $t$, is \begin{equation}
    \Delta r(t) \approx \frac{\varphi^2R}{2} \frac{t}{\tau} \left( 1 - \frac{t}{\tau} \right),
    \label{eq:delta_r_angle}
\end{equation}
where $\varphi$ is the fraction of an orbit subtended over \gaia's lifetime $\tau$. Figure \ref{fig:detectability_diagram} shows this quantity. Since $\varphi \approx V \tau / R$, where $V$ is the orbital velocity (assumed to be constant throughout \gaia's lifetime), we can express $\Delta r$ in Equation \ref{eq:delta_r_angle} as:
\begin{equation}
    \Delta r(t) \approx \frac{V^2 t}{2 R} \left( \tau - t \right).
\end{equation}

Since \gaia\ detects the angular position of a star at a distance $D$ with a precision of $\sigma_G$ rather than a physical distance, the significance of any individual observation, $\sigma(t) = \Delta r(t) / D\sigma_G$ can be represented as:
\begin{equation}
    \sigma(t) \approx \frac{V^2}{2 R D} \frac{1}{\sigma_G} t \left( \tau - t \right).
\end{equation}
The overall significance of a series of $N$ independent observations is the quadrature sum of the individual significances, $\sigma^2 = \sum\ \sigma^2 (t)$. We can then estimate the significance of a detection of orbital acceleration by equally spaced observations, in which case:
\begin{eqnarray}
\sigma &\approx& \frac{V^2}{2 R D} \frac{\tau^2}{\sigma_G} \left[\sum_{i=0}^N\ \left(\frac{i}{N}\right)^2 \left( 1 - \frac{i}{N} \right)^2 \right]^{1/2} \\
&\approx& \frac{V^2}{2 R D} \frac{\tau^2}{\sigma_G} \sqrt{\frac{N}{30}} \label{eq:sigma_approx} \\
&\approx& 14 \left(\frac{V}{5\ {\rm km\ s}^{-1}}\right)^2 \left(\frac{R}{10^3\ {\rm AU}}\right)^{-1} \left(\frac{D}{100\ {\rm pc}}\right)^{-1} \nonumber \\
& & \quad \times \left(\frac{\tau}{5\ {\rm yr}}\right)^2 \left(\frac{\sigma_G}{10\ \mu{\rm as}}\right)^{-1} \left(\frac{N}{70}\right)^{1/2}. 
\label{eq:detectability_sigma}
\end{eqnarray}
The approximation for the sum in Equation \ref{eq:sigma_approx} is true in the limit of large $N$, a reasonable approximation since \gaia\ observes a typical star $\simeq$70 times. Using typical values for a nearby star, Equation \ref{eq:detectability_sigma} suggests that \gaia\ could detect orbital acceleration for binaries with separations as large as 10$^3$ AU, with a signal-noise ratio $\gtrsim$10.

However, Equation \ref{eq:detectability_sigma} is only an imperfect approximation for \gaia's detectability. A realistic detection limit must additionally account for the simultaneous observability of the star's parallax (or distance) and proper motion, both of which degrade with distance. To test this limit, we run a series of simulations, placing fake 1 $\Msun$ stars in orbits of various velocities and curvature radii at distances of 10 pc, 100 pc, and 1 kpc. For each star on its orbit, we then take mock observations representative of a realistic \gaia\ cadence and precision\footnote{\gaia's astrometric precision, $\sigma_G$, is the precision with which \gaia\ can measure a star's position in the along-scan direction, a quantity that depends on a star's magnitude. Brighter stars have improved astrometric precision, but saturate at $G<12$. In calculating results throughout this work, we use our own fits to $\sigma_G$ based on the theoretical astrometric precision of \gaia\ as a function of magnitude, as provided in \citet{Gaia_astrometry}. The best-measured, bright stars can reach $\sigma_G\simeq17$ mas.} following the procedure defined in Paper I, but for only the partial orbits considered here. We refer the reader to Paper I for details of our procedure. Here, we only comment that for the set of observations of each star over \gaia's lifetime, we run a Markov chain Monte Carlo sampler \citep[we use {\tt emcee};][]{Foreman-Mackey2013} to determine constraints on the derived orbital acceleration.

\begin{figure*}
    \begin{center}
    \includegraphics[width=1.0\textwidth]{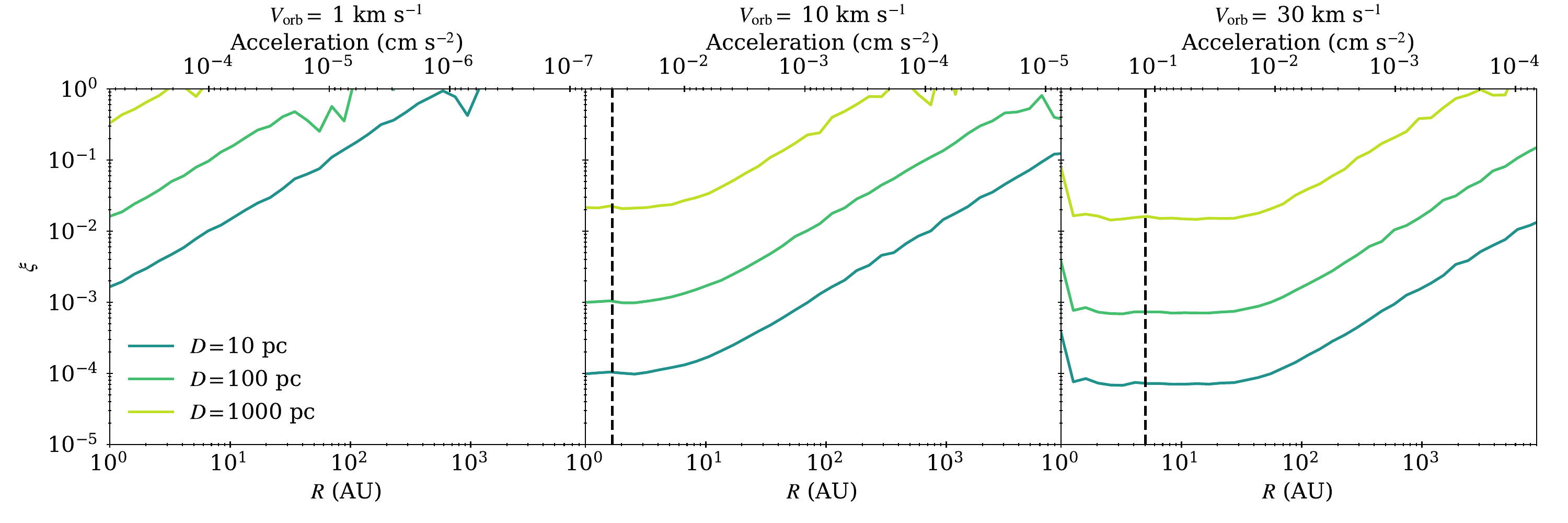}
    \caption{ We calculate \gaia's ability to detect orbital accelerations in a 1 \Msun\ star for different orbital velocities and radii of curvatures. A value of $10^{-1}$ for $\xi$ implies a 10\% measurement accuracy on the acceleration, so smaller values on the ordinate axis correspond to more precise measurements. These results follow the expected trends from Equation \ref{eq:detectability_sigma}: increased accelerations, either through shorter orbital periods or more massive dark companions, improve the measurement ability, as does smaller distances from the Sun. For long period orbits, where orbital velocities may be $\sim$1-10 km s$^{-1}$, accelerations may be detectable for binaries at $\sim$100 pc with orbital separations $\sim$100 AU. For more massive, brighter stars, \gaia\ may be sensitive to astrometric binaries at even farther distances.}
    \label{fig:model_results}
    \end{center}
\end{figure*}

Figure \ref{fig:model_results} shows the resulting constraints for three different sampled velocities. Several trends are worth explicitly stating. First, the detectability $\xi$ (which we define as the fractional uncertainty on the acceleration measurement) scales inversely with distance (as expected from Equation \ref{eq:detectability_sigma}), so the best limits will be on stars nearest to the Sun. Second, systems with larger velocities - and therefore larger accelerations - are better detected. Third, detectability generally decreases with radius of curvature (again, as expected from Equation \ref{eq:detectability_sigma}). However, for sufficiently small $R$, our procedure breaks down, since \gaia\ starts to observe an appreciable fraction of the orbit; a simple arc no longer accurately describes the motion of a star. This regime appears as a saturation in the measurement precision in Figure \ref{fig:model_results}, which occurs roughly at the point where \gaia\ observes one-tenth of an orbit over its lifetime. The vertical, dashed lines in the second and third panels show the orbital separation where a binary makes a complete orbit over \gaia's lifetime. 

It is worth noting here that by assuming a constant acceleration over \gaia's lifetime our procedure makes a significant approximation. For orbits with periods longer than $\simeq10$ times \gaia's lifetime, our approximation is reasonable. However, for orbits with periods between one and ten times \gaia's lifetime, one can start to fit orbital solutions to the evolving accelerations exhibited by a star in an astrometric binary. As we show in Paper I, complete orbital solutions are not accurately derivable for binaries in periods longer than \gaia's lifetime; however in this transition region, we suggest that our observations provide a lower limit to the precision with which an astrometric orbit can be characterized. A more complete quantitative analysis of what can be derived from observations of stars with evolving accelerations is outside the scope of this work.

Using the results from our simulations assuming a constant acceleration, we can empirically determine \gaia's detection precision for astrometric orbits:
\begin{equation}
    \xi = \frac{\sigma (\Delta V^2/R)}{\Delta V^2/R} \approx 35 \frac{D R}{V^2} \frac{\sigma_G}{\tau_G^2} \sqrt{\frac{1}{N}}
    \label{eq:detectability_empirical}.
\end{equation}
This empirical result is within a factor of three of our approximation provided in Equation \ref{eq:detectability_sigma}.

The constraints so far are derived under the assumption of face-on orbits, but of course the orbits are oriented isotropically with respect to the line of sight from Earth. Using the Campbell elements, orbital orientations are quantified using three angles: the inclination, the longitude of the ascending node, and the argument of periapse. In general it is impossible to fully derive the orbital orientation in three dimensions for the systems discussed here, since only a fraction of the orbit is observed. The degeneracy between these three orientation angles for an individual system simplifies the problem considerably.

\begin{figure}
    \begin{center}
    \includegraphics[width=1.0\columnwidth]{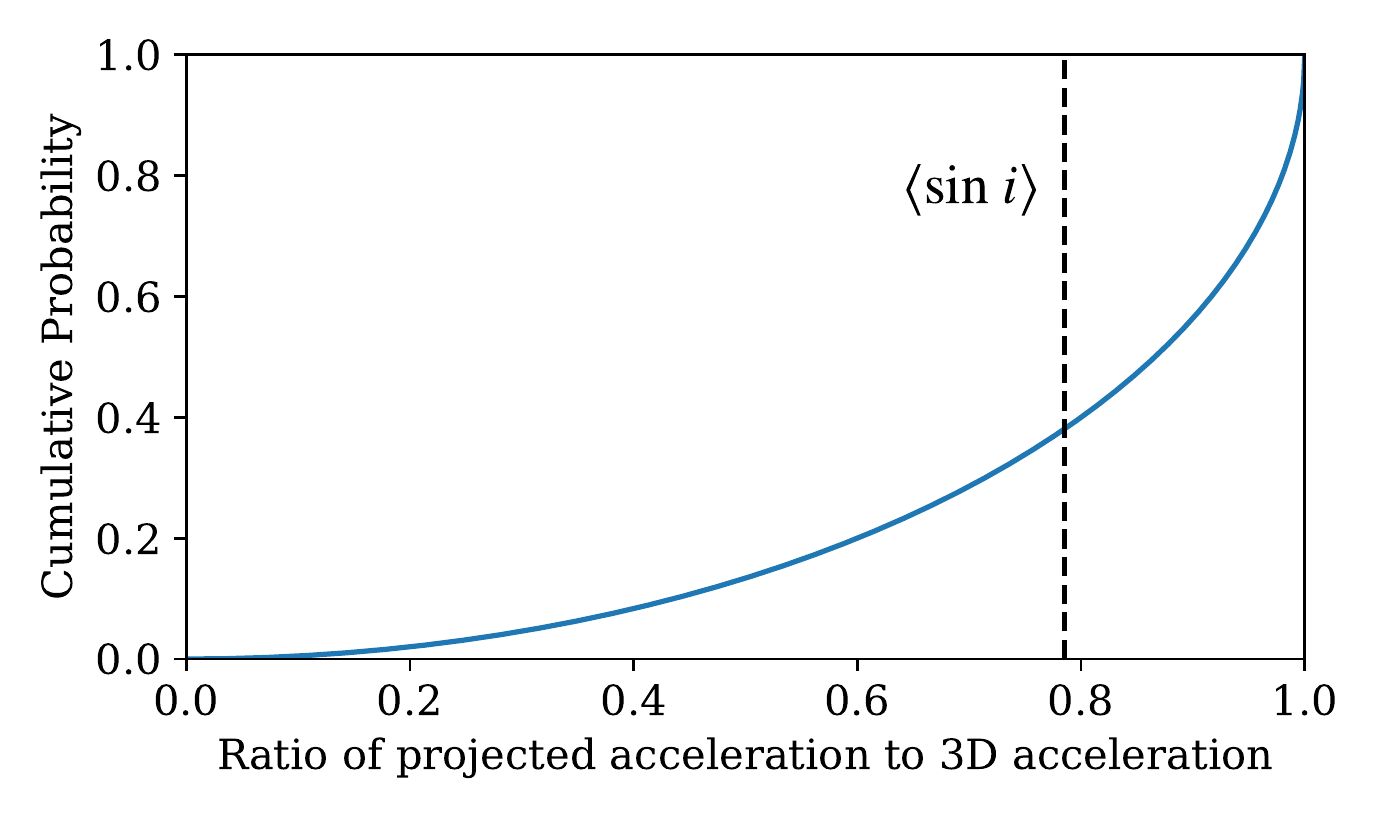}
    \caption{ The 3D acceleration vector is measured by the 2D projection onto the proper motion plane. For a distribution of randomly oriented systems, we show the cumulative distribution of the ratios between the magnitude of the 2D projected vectors with the full 3D vectors. Most ratios are close to unity, with a mean ratio of $\simeq$0.8.}  
    \label{fig:inclination}
    \end{center}
\end{figure}

The orbital acceleration on a star from an unseen companion can be quantified as a vector, which results in a change in the star's velocity. Observationally, this can be measured as a change in the two dimensions of proper motion (in $\alpha$ and $\delta$) and in the radial velocity. In principle, through careful observations, a change in the radial velocity of a star over years to decades can be measured\footnote{For sufficiently bright ($V\lesssim13$) stars the RVS instrument on \gaia\ can measure radial velocities with a precision of $\lesssim$1 km s$^{-1}$. Over a nominal mission lifetime of 5 years, \gaia\ can therefore achieve a measurement of the radial velocity component of the acceleration vector as small as (1 km s$^{-1}$/5 years) $\sim 10^{-4}$ cm s$^{-2}$. Comparison with Figure \ref{fig:model_results} shows this acceleration precision rivals the astrometric measurement precision on acceleration. As the radial velocity measurement precision depends on both the type of star being observed as well as its magnitude, we ignore it in this work.}; however, here we will assume the acceleration is only measured astrometrically through a change in the proper motion over time. Our observations are therefore limited to the projection of the acceleration of the 3D velocity onto the 2D proper motion plane. Figure \ref{fig:inclination} compares the magnitude of the 3D accelerations to that of the 2D projected accelerations for a distribution of randomly oriented systems. The cumulative distribution of this ratio shows that most binaries have a ratio close to unity, with mean ratio $\left<\sin i\right> \simeq 0.8$. For a population of randomly oriented orbits, like those observed by \gaia, this factor is almost negligible, especially compared with the other uncertainties in the analysis such as the unknown distribution of companion masses, orbital separations, and eccentricities. 

For a pathologically oriented system, viewed with the acceleration vector along the line of sight, the component in the plane of the celestial sphere is zero. However, this is only for a very small subset of orbits; in most cases, the projected acceleration will be very similar to the full 3D acceleration.

\section{Elliptical Orbits Revisited}
\label{sec:theory}

Now that we have an estimate for \gaia's sensitivity to orbital accelerations, we wish to translate that limit into orbital parameters. For circular orbits with a luminous star of mass $M_1$ and a dark star of mass $M_2$, this is trivial; since the radius of curvature is the orbital separation $a$, and the orbital velocity $V$ can be determined as $V^2 = \mathcal{G}(M_1+M_2)/a$, when accounting for the translation into a barycentric reference frame, we find the constant magnitude of the acceleration on the luminous star:
\begin{equation}
    \left. \frac{V^2}{R}\right|_1 = \frac{\mathcal{G} M_2}{a^2}. \label{eq:accel_circular}
\end{equation}
The subscript on the l.h.s.\ is to clarify that the acceleration is felt by star 1. 

\begin{figure}
    \begin{center}
    \includegraphics[width=1.0\columnwidth]{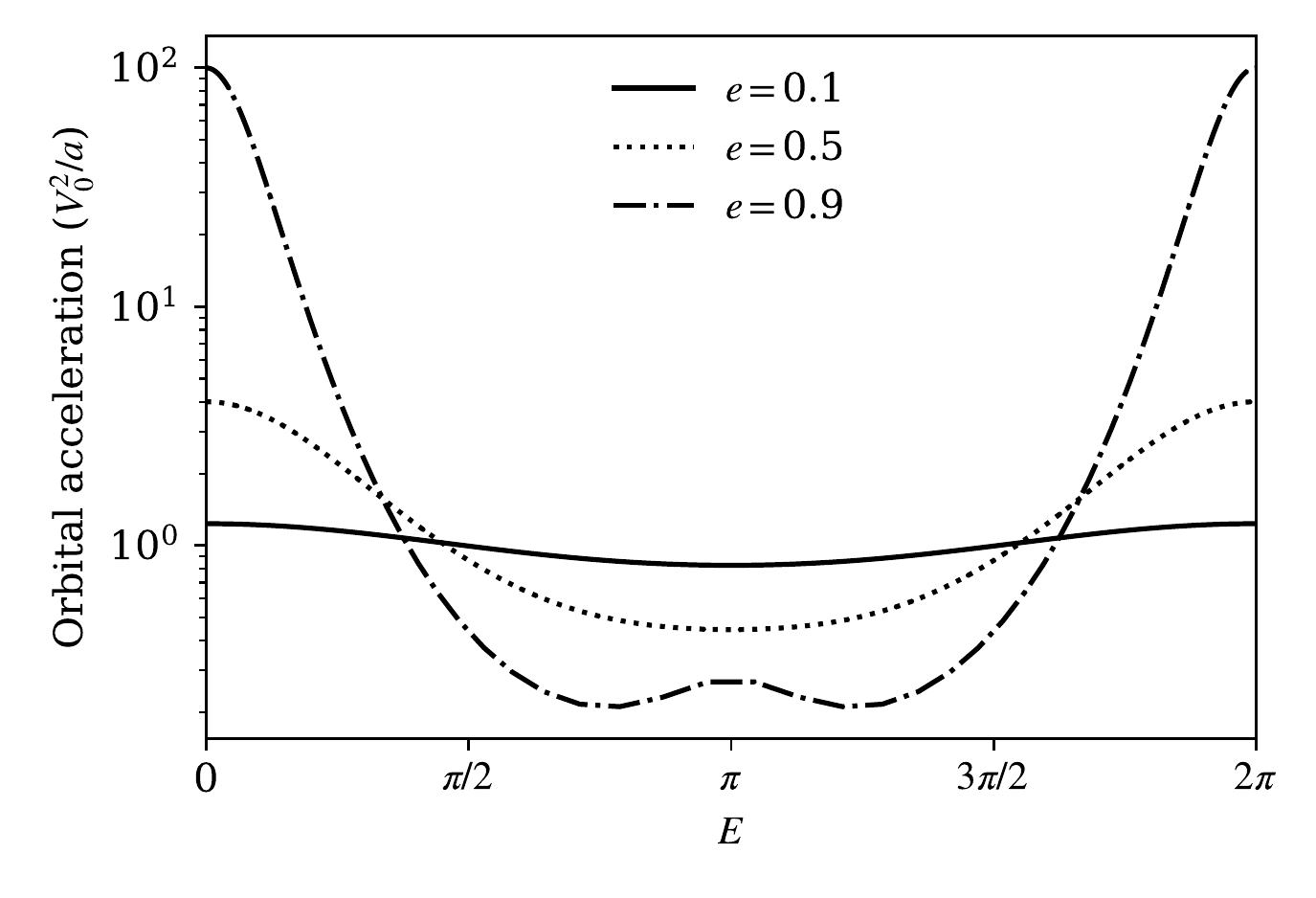}
    \caption{ The acceleration of an orbit as a function of eccentric anomaly relative to a circular orbit. Accelerations are larger (smaller) near periastron (apastron). For orbits with $e>2/3$, there is a double minima near the orbital apocenter.}  
    \label{fig:elliptical_orbital_acceleration}
    \end{center}
\end{figure}

When accounting for eccentric orbits, the orbital acceleration varies as a function of orbital phase, quantified by the eccentric anomaly, $E$. The radius of curvature for an eccentric orbit varies with $E$ as
\begin{equation}
    R = \frac{a}{\sqrt{1-e^2}} \left( 1 - e^2 \cos^2 E\right)^{3/2}.
\end{equation}
At the same time, the orbital velocity for an eccentric orbit varies as
\begin{equation}
    V^2 = \frac{\mathcal{G}(M_1+M_2)}{a} \left( \frac{1 + e \cos E}{1 - e \cos E} \right).
\end{equation}
Combining the two and moving to the barycentric frame, we find the orbital acceleration for the luminous star:
\begin{equation}
    \left. \frac{V^2}{R} \right|_1 = \frac{\mathcal{G}M_2}{a^2} \left( \frac{1 + e \cos E}{1 - e \cos E} \right) \frac{\sqrt{1-e^2}}{\left( 1 - e^2 \cos^2 E \right)^{3/2}}.
    \label{eq:elliptical_orbital_acceleration}
\end{equation}
Figure \ref{fig:elliptical_orbital_acceleration} shows the orbital acceleration as a function of orbital phase (represented by the eccentric anomaly) for three separate orbital eccentricities. These are each normalized by $V_0^2/a = \mathcal{G} M_2/a^2$, so a value of unity corresponds to the circular equivalent acceleration. Low eccentricity orbits have accelerations very close to $V_0^2/a$ whereas the orbital accelerations increase (decrease) around periastron (apastron) for larger eccentricities. Orbits with $e>2/3$ exhibit a double-minima feature around apastron, as shown by the dot-dashed line in Figure \ref{fig:elliptical_orbital_acceleration}.

\begin{figure}
    \begin{center}
    \includegraphics[width=1.0\columnwidth]{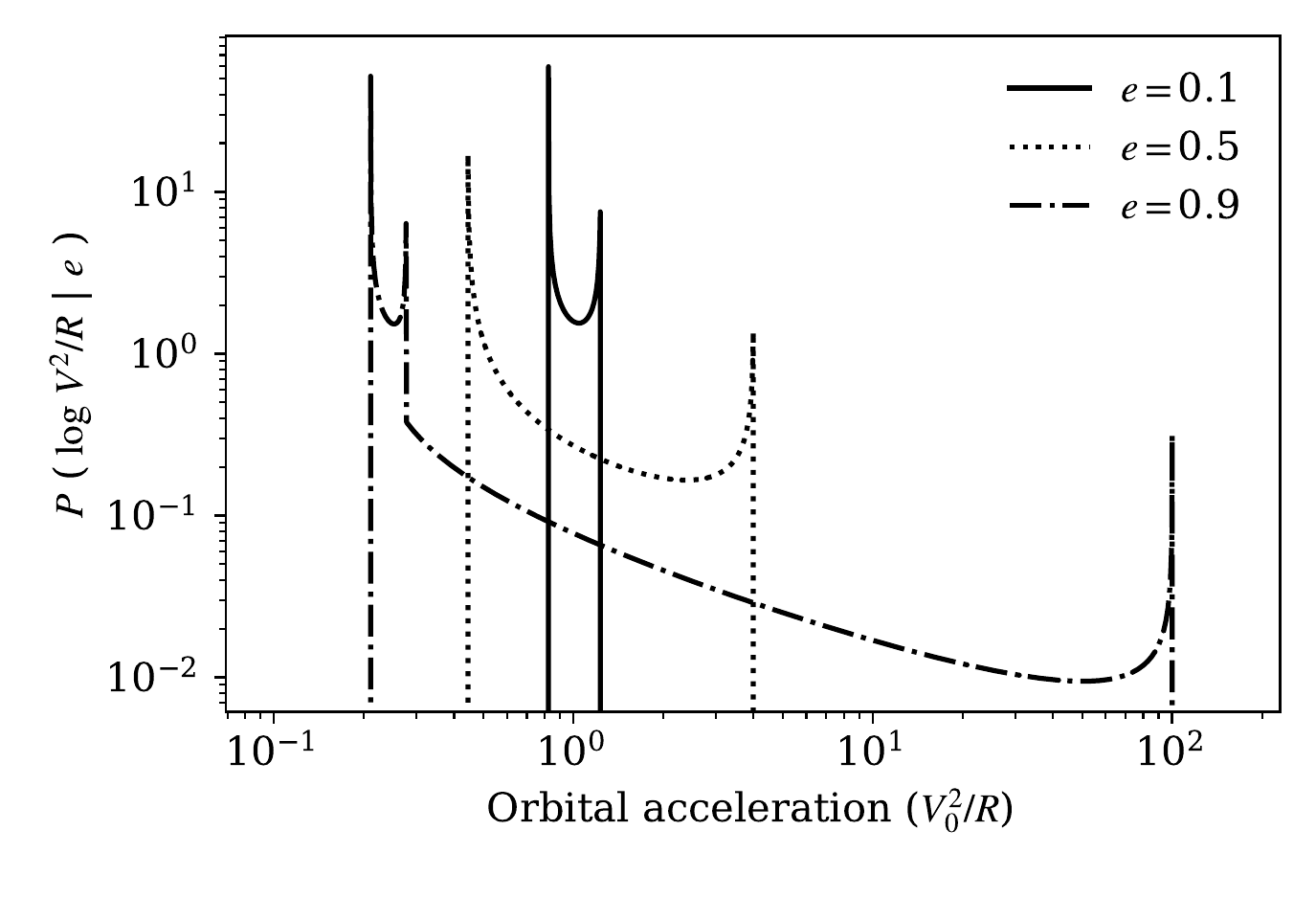}
    \caption{ Convolving the acceleration felt by a star with the length of time it spends at each part of an orbit provides the likelihood of observing a star at different orbital accelerations. For low eccentricity orbits, the circular acceleration is a reasonable representation. In contrast, highly eccentric orbits are most likely to be observed with accelerations a factor of a few less than the circular approximation. }  
    \label{fig:eccentricity_contribution}
    \end{center}
\end{figure}

Since binaries spend more time near apastron than near periastron, Equation \ref{eq:elliptical_orbital_acceleration} must be combined with a function quantifying the amount of time spent at each orbital phase, $P(E) = 1-e\cos E$, to determine the typical orbital accelerations observed for a population of binaries. After combining, we find:
\begin{eqnarray}
    P\left[ \left. (V^2/R) \right| e \right]_1 = \frac{P(E)} {\sum_{i=1}^N \left| \partial (V^2/R)_1 / \partial E \right|_{E=E^*}},
    \label{eq:prob_elliptical_orbital_acceleration}
\end{eqnarray}
where the summation is over the $N$ separate eccentric anomalies, $E^*$, between 0 and $2\pi$ corresponding to a given combination of $V^2/R$, $V_0^2/a$, and $e$. For orbits with $e<2/3$, Figure \ref{fig:elliptical_orbital_acceleration} shows that two solutions are possible, while as many as four solutions are available for orbits with $e>2/3$. A closed-form solution does not exist for $E^*$, so Equation \ref{eq:prob_elliptical_orbital_acceleration} must be solved numerically. Figure \ref{fig:eccentricity_contribution} shows this corresponding probability, again normalized by $V_0^2/a$. As expected, for low eccentricities, orbital accelerations are typically found near the circular values; even for orbits with $e=0.5$, orbital accelerations are within a factor of a few of their circular equivalents. However, since eccentric binaries spend more time near apastron, orbits with larger eccentricities have orbital accelerations somewhat smaller than their circular values. We note that the spikes in Figure \ref{fig:eccentricity_contribution} are caused by the  $\left| \partial (V^2/R)_1 / \partial E \right|$ term in the denominator of Equation \ref{eq:prob_elliptical_orbital_acceleration} going to zero at periastron and apastron.

\begin{figure}
    \begin{center}
    \includegraphics[width=1.0\columnwidth]{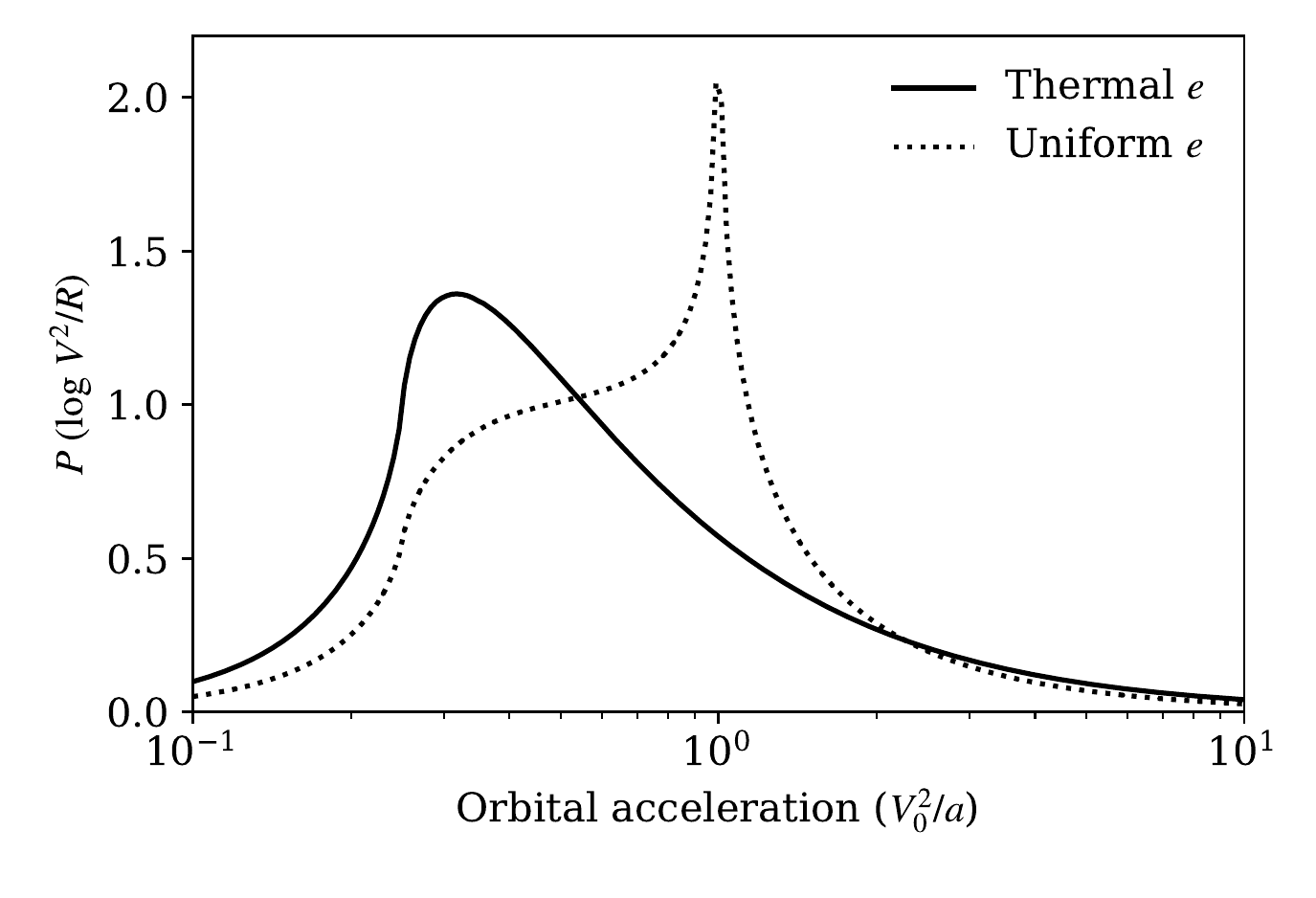}
    \caption{ If convolved with a distribution of orbital eccentricities, we can determine the distribution of orbital accelerations expected. For a uniform distribution of eccentricities, we expect most binaries to have eccentricities near the circular value, $V_0^2/a$, whereas for a thermal eccentricity distribution, most orbits have accelerations a factor of $\approx$3 smaller than the circular value. }  
    \label{fig:eccentricity_distribution}
    \end{center}
\end{figure}

By convolving Equation \ref{eq:prob_elliptical_orbital_acceleration} with a distribution of eccentricities, $P(e)$, we can obtain the typical orbital accelerations expected from observing a ``snapshot" of a population of binaries, again expressed in terms of $V_0^2/a$:
\begin{equation}
     P\left( V^2/R \right)_1 = \int_0^1 {\rm d}e\  P\left[ (V^2/R) | e \right]_1\ P(e). 
\end{equation}
Figure \ref{fig:eccentricity_distribution} shows the resulting likelihoods of detecting orbital accelerations, normalized by $V_0^2/a$, for two separate eccentricity distributions. For a uniform eccentricity distribution, $P(e)=1$, the distribution of orbital accelerations peaks strongly around the circular value, but with a significant tail with smaller values due to eccentric binaries being found near apocenter ($\left< V^2/R \right> = 0.46$). However, if we adopt a thermal eccentricity distribution, $P(e)=2e$, the increased preference for binaries with higher eccentricities pushes the overall distribution toward smaller orbital accelerations. The resulting distribution peaks at $\simeq 1/3$ the circular value ($\left< V^2/R \right> = 0.41$). 

\section{The Compact Object Population}
\label{sec:pop_synth}

\subsection{Population Synthesis Predictions}

\begin{figure}
    \centering
    \includegraphics[width=1.0\columnwidth]{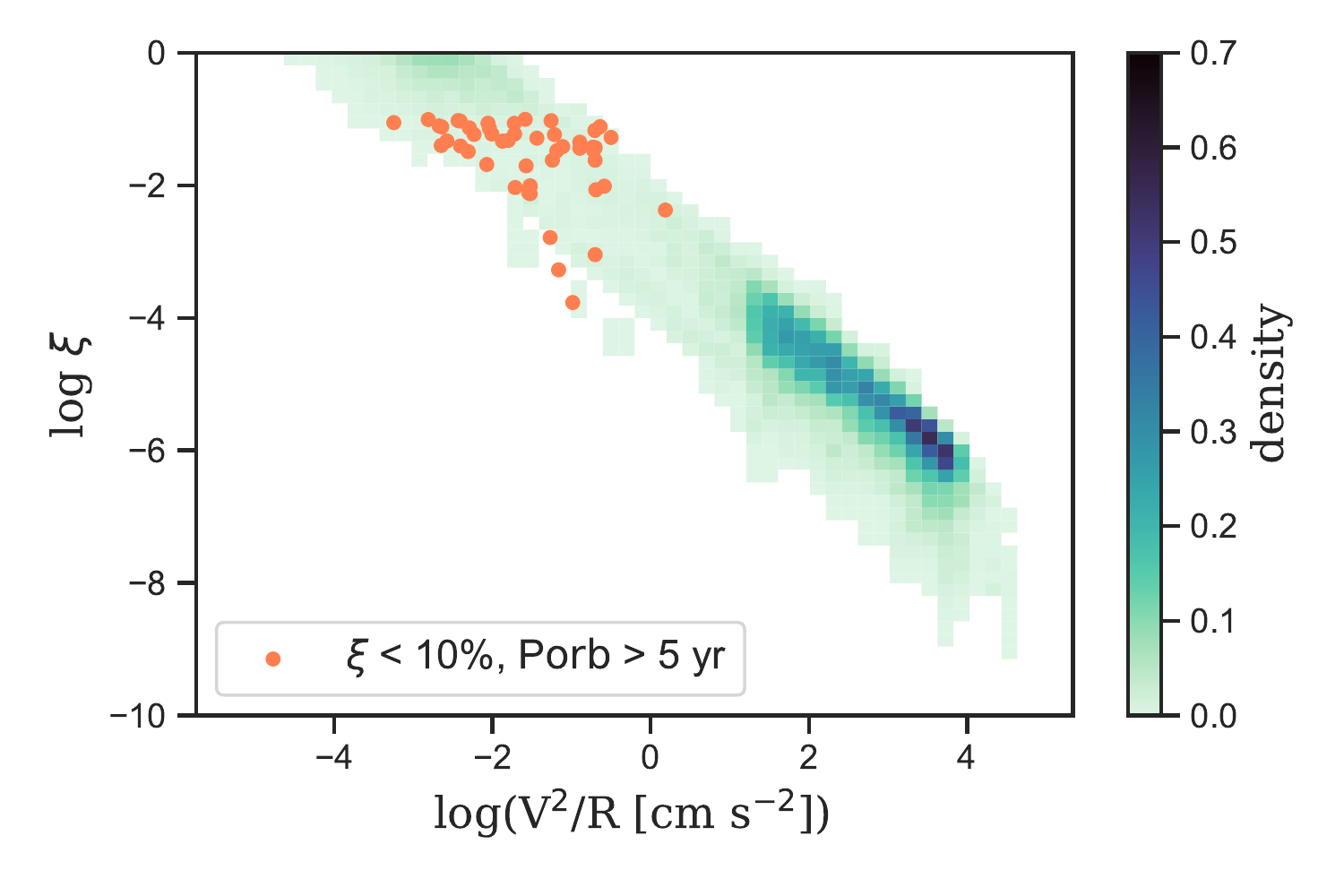}
    \caption{Contours indicate the distribution of $V^2/R$ and $\xi$ (both are provided as a base-10 logarithm), \gaia's measurement precision on that acceleration, for a the population of BH binaries in the Milky Way. The bulk of systems, those with large $V^2/R$, have $P_{\rm orb} < 5$ yr, and are the subject of Paper I. Here, we focus on those systems that show incomplete orbits, with $\Porb \lesssim$5 years. Orange points show the binaries produced by one galactic realization that satisfy our observability constraints.}
    \label{fig:cosmic}
\end{figure}

We obtain an expectation for the number of BHs to be astrometrically observed by \gaia\ using binary population synthesis. For a complete description of the population generation parameters and astrophysical implications -- including predictions for observability of astrometric binaries by {\it Gaia} -- see Chawla et al.~(in prep). Here, we describe the basic procedure, which is based on that presented in \citet{Breivik2018}. We start by generating a population of initial stellar binaries with primary masses distributed according to \citet{Kroupa2001} and secondary masses distributed uniformly with a binary fraction of $0.5$. We assign each binary with an initial separation from a log-uniform distribution between a minimum value such that the system does not begin in Roche overflow and up to $10^5\,\rm{R}_{\odot}$ \citep{Han1998}. Finally, we assume that eccentricities are initially thermally distributed \citep{Heggie1975}.

While a complete description of the default parameters in {\tt COSMIC} is provided by \citep{breivik2020}, we highlight the most significant model prescriptions related to mass transfer and core collapse. Mass transfer stability is determined using the mass ratio conditions described in \citep{belczynski2008}, with stable mass transfer following the procedure described in \citet{hurley2002}. We evolve binaries going through a common envelope using the $\alpha\lambda$ prescription, where we set $\alpha=1$, corresponding to an efficient envelope ejection and $\lambda$ is determined using the prescription described in the Appendix of \citet{claeys2014}. For stars going through core collapse, we follow the {\tt delayed} prescription from \citet{fryer2012}. This prescription determines both the BH masses, as well as the amount of supernova ejecta that is accreted by the newly born BH as fallback mass. While BHs are initially given natal kicks drawn from a Maxwellian with $\sigma=265$ km s$^{-1}$, most BHs receive kicks close to or equal to zero, as fallback proportionally reduces the strength of the natal kick.

Those binaries that become BHs with luminous companions are then randomly sampled multiple times in its orbit at different eccentric anomalies and placing them in different locations throughout the Milky Way so the rates match the star formation rate, metallicity, and stellar density of the m12i galaxy in the Latte suite of the FIRE-2 simulation \citep{Wetzel2016,hopkins2018}.

Finally, for each binary we calculate the apparent acceleration of the luminous object following Equation \ref{eq:elliptical_orbital_acceleration}. We then sample 50 random Milky Way iterations to take into account the effects from stochastic sampling. Our procedure additionally includes two major improvements over the populations presented in \citet{Breivik2018}: whereas \citet{Breivik2018} focused on giant stars, we account for all types of luminous companions, and we further include a dust model to account for extinction depending on each synthetic star's position in the Milky Way.

From our sample of synthetic binaries, Figure \ref{fig:cosmic} compares the distribution of orbital accelerations with the measurement precision of \gaia\ on those accelerations. Contours are generated from 50 random Milky Way iterations and show the distribution of all systems comprised of a BH and a luminous star. Our simulations predict a large sample of binaries with $\log\ V^2/R>1$; these systems have orbital periods much less than 5 yrs and are therefore the subject of Paper I (the value of $\xi$ derived from Equation \ref{eq:detectability_empirical} is an extrapolation). The results in this work are relevant for systems with binary periods $\gtrsim$5 years. After restricting for systems with an acceleration precision $<$0.1, $P_{\rm orb}>$5 years, and a \gaia\ $G$ magnitude $<$20, we find \gaia\ ought to detect $\simeq$47 such systems. We highlight in Figure \ref{fig:cosmic} the systems generated by one particular Milky Way realization that satisfy these observability constraints (orange markers). While our $\simeq$47 expected systems are derived from 50 random Milky Way realizations, we find 31 detectable systems in this particular example.

\begin{figure}
    \centering
    \includegraphics[width=1.0\columnwidth]{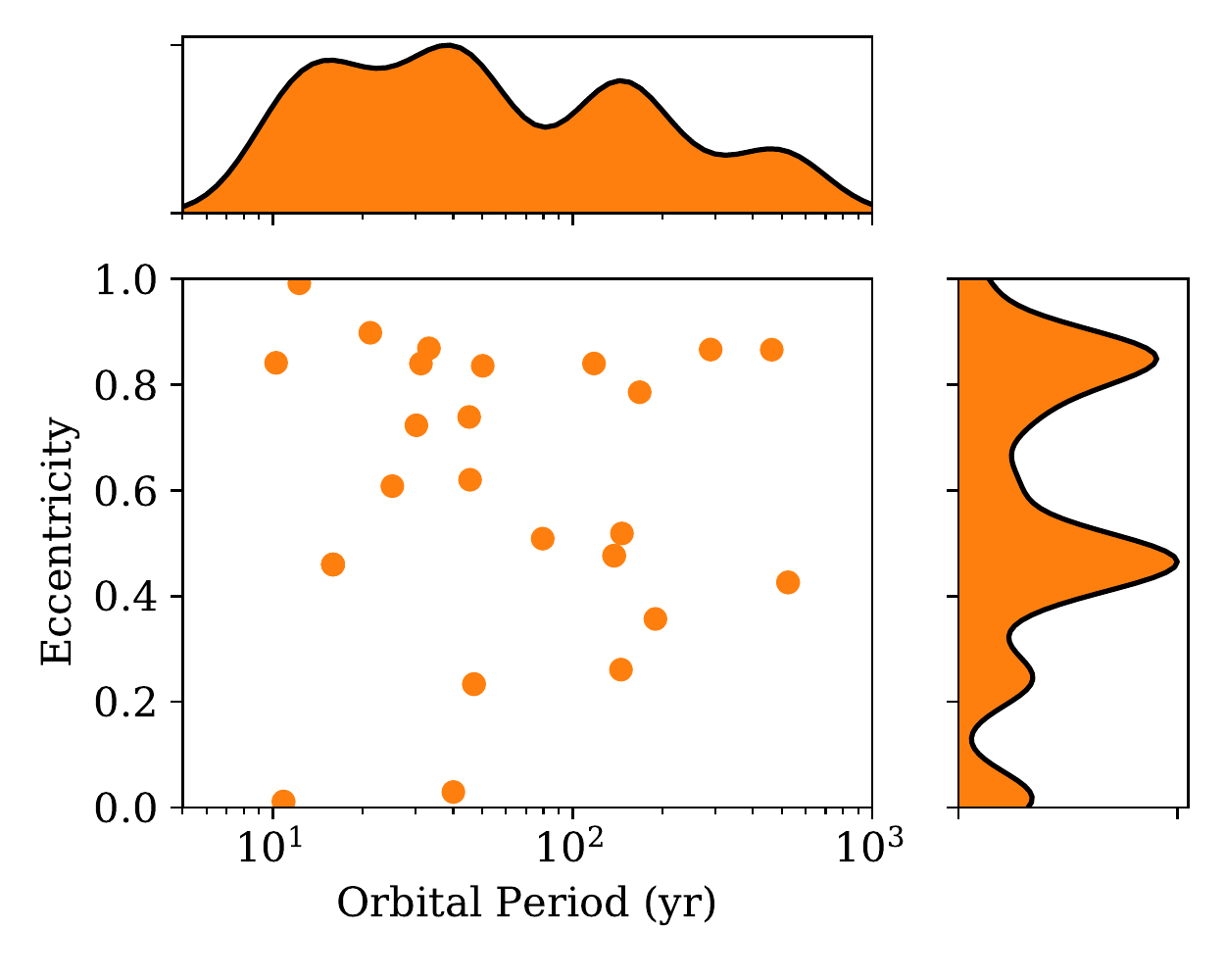}
    \caption{The orbital characteristics of binaries generated for one Milky Way realization that show accelerations detectable by \gaia. We have selected for binaries with $\porb>10$ yr and $\xi<10$\%. Our models predict a range of orbital eccentricities, and a preference for shorter orbital periods. }
    \label{fig:cosmic_orbits}
\end{figure}

In Figure \ref{fig:cosmic_orbits}, we provide the log $\porb$ and $e$ for the same example Milky Way realization. The top and right panels provide one-dimensional histograms of log $\porb$ and $e$, from which we see there is a preference for shorter $\porb$ systems and a slight preference for higher $e$ systems. Importantly, systems span the entire eccentricity range, and there is no clear cutoff at large $\porb$. This suggest that a future instrument even more sensitive than \gaia\ would be able to detect systems at even larger $\porb$.

\begin{figure}
    \centering
    \includegraphics[width=1.0\columnwidth]{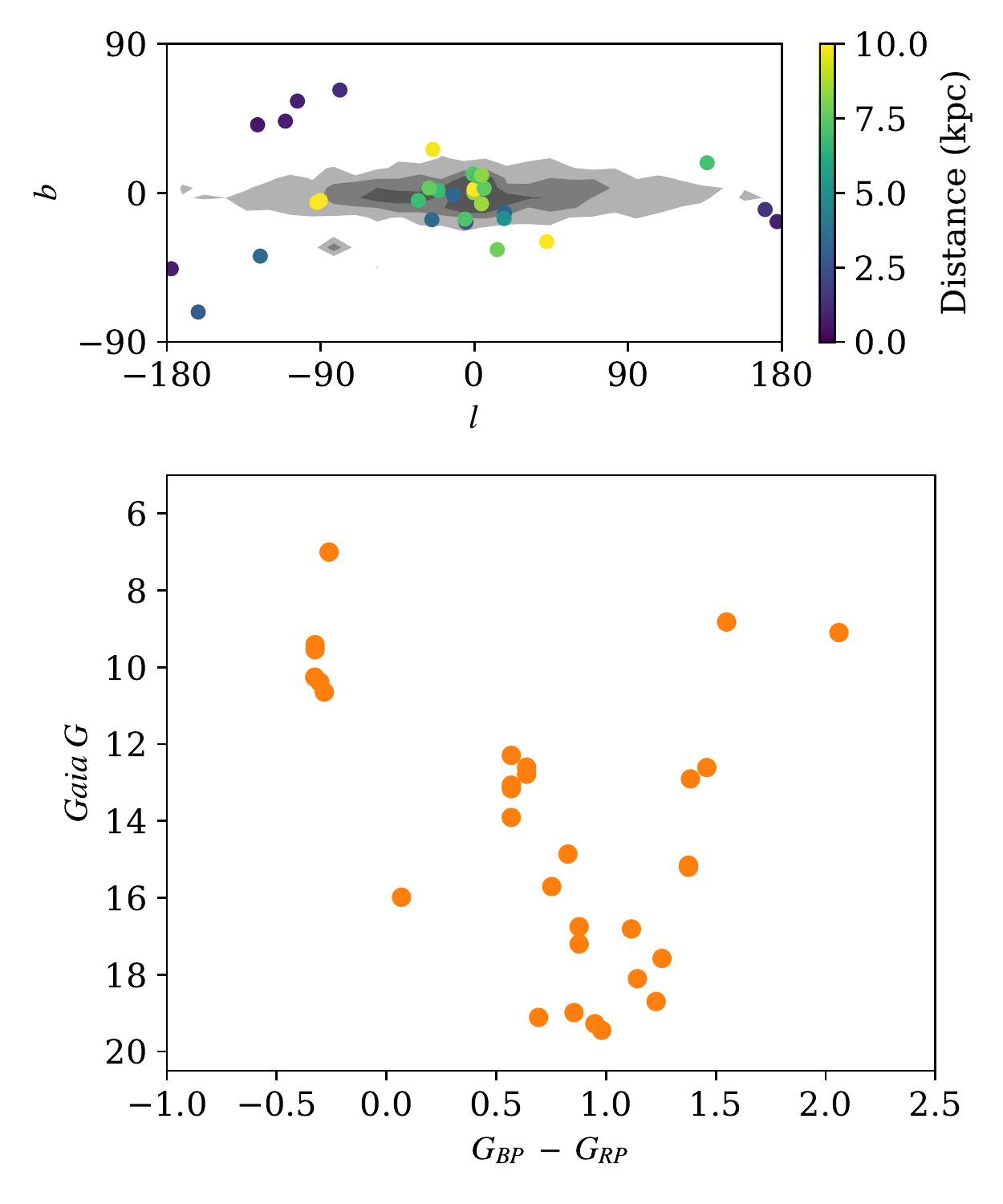}
    \caption{ For our sample Milky Way realization, we show the Galactic positions and distances of our synthetic binaries (top panel) and \gaia\ $G$ magnitude and $BP-RP$ color (bottom panel). Grayscale contours in the top panel indicate the positions of random \gaia\ EDR3 stars. The systems found at low Galactic latitudes, near the center of the Milky Way are typically found at large distances ($\gtrsim 1$ kpc), suggesting crowding may make some of these systems difficult to detect. However, a number of systems are expected to exist nearby in low-density regions of the sky. The bottom panel shows that the luminous stars in these binaries span a range of colors and magnitudes, with the most luminous objects accessible to follow-up campaigns with even modest-sized telescopes.}
    \label{fig:cosmic_characteristics}
\end{figure}

In Figure \ref{fig:cosmic_characteristics}, we further describe the observational characteristics of the binaries in our synthetic Milky Way realization. In the top panel, we show the Galactic coordinates of these binaries (circular markers), with the marker colors indicating the distance to each system. A large fraction of systems are found near the center of the Milky Way, at distances of several kpc. Grayscale contours indicate the positions of a random sample of \gaia\ EDR3 stars, suggesting that many of these sources will be found in crowded fields. However, a number of nearby ($\lesssim$1 kpc) systems exist far from the Milky Way center. The bottom panel shows that the luminous companions to the BHs in our sample of binaries span a wide range of magnitudes and \gaia\ colors. The brightest of these are accessible to even relatively small science-grade telescopes for observational follow-up.

\subsection{Uniquely Identifying Compact Object Companions}

Although our simulations suggest that \gaia\ is sensitive to $\simeq$47 long-period binaries with BH companions, degeneracies prevent some binaries from being uniquely identified; the small acceleration produced by a massive BH companion in a long-period orbit, can also be caused by a low-mass, late-type star in somewhat shorter-period orbit. From Equation \ref{eq:accel_circular} the degeneracy can be expressed as $M_2/a^2 = {\rm constant}$. 

How then to uniquely identify systems containing BH companions? In many cases such identification may not be possible. For binaries sufficiently close-by, follow-up observations with adaptive optics techniques may be able to resolve faint, low-mass companions in wide orbits. Yet even with the non-detection of a companion using state-of-the-art techniques and large aperture telescopes, a ``lighter" companion in an even smaller orbit could also account for the observed acceleration. However, there is a mass limit below which the orbital period is too small to explain the observed acceleration; for small companions with sufficiently short orbital periods, \gaia\ ought to see a complete orbit over its lifetime (or at least measure a higher-order ``jerk" term in the star's motion). The fact that \gaia\ observes only part of an orbit sets a lower limit on the mass of a putative companion. Combining Equation \ref{eq:accel_circular} with the generalized form of Kepler's Third Law, and replacing $P_{\rm orb}$ with twice \gaia's lifetime (we assume an upper limit on the orbital period can be placed if at least half the orbit is observed) we find a relation between the two stars' masses and the acceleration of the luminous star:
\begin{equation}
     1 = \frac{\pi^4}{\tau_{\rm Gaia}^4} \frac{\mathcal{G}M_2^3}{(M_1+M_2)^2} \left( \left. \frac{V^2}{R} \right|_1 \right)^{-3}.
\label{eq:mass_constraint}
\end{equation}
Note that we have assumed a circular orbit in deriving Equation \ref{eq:mass_constraint}, a point we address at the end of this section.

Combining the mass constraints from Equation \ref{eq:mass_constraint} with high-angular-resolution follow-up imaging of any star showing large accelerations allow for the compact object nature of putative companions to be confirmed. Even faint white dwarf companions can be identified using this method. 

\begin{figure}
    \begin{center}
    \includegraphics[width=1.0\columnwidth]{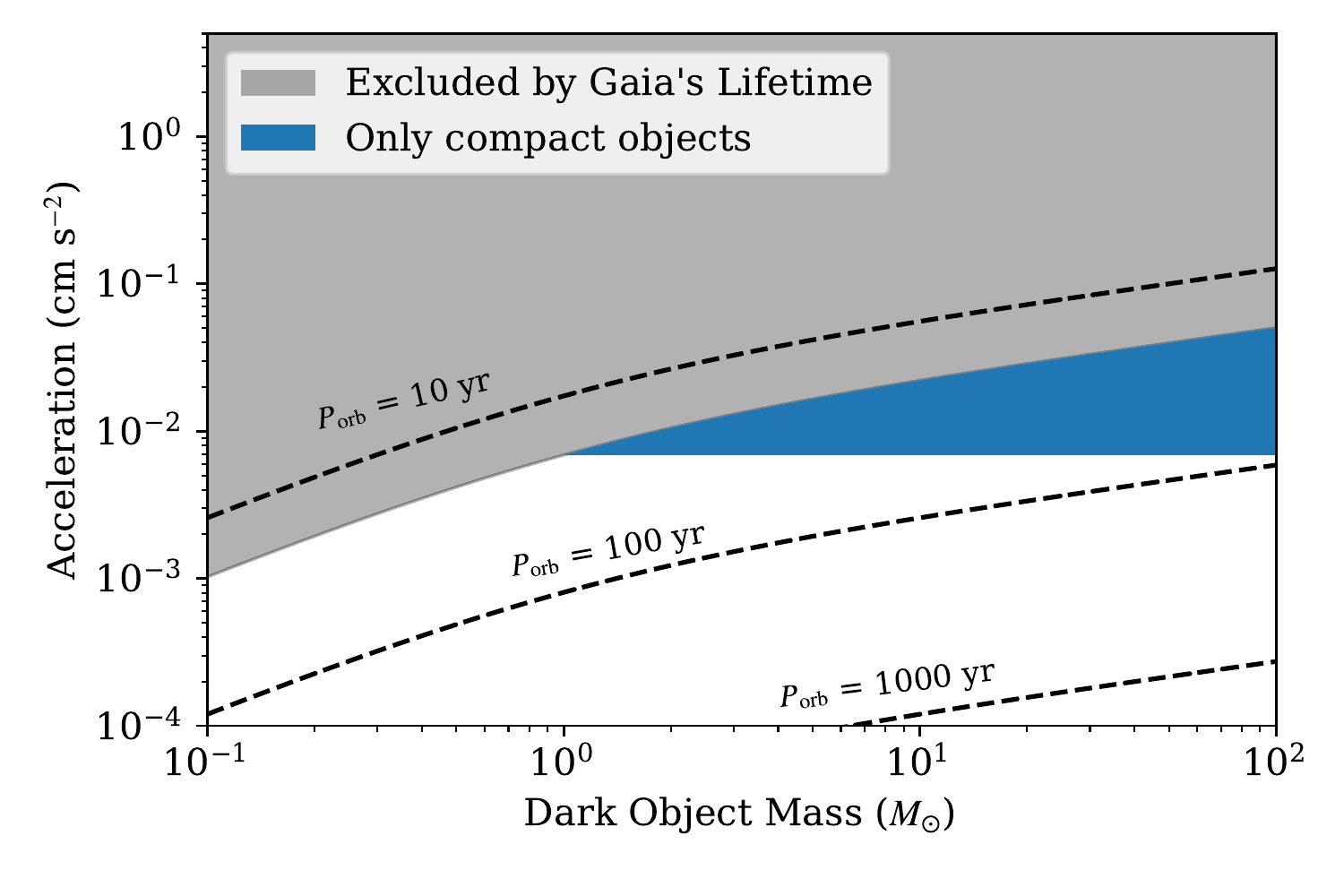}
    \caption{ The gravitational acceleration on a luminous star depends on the distance and mass of a companion object. BHs in long period orbits can be confused with a faint, low-mass star on a short period orbit. However, for sufficiently large accelerations, a low-mass stellar companion would have to be in an orbit with a short enough period that \gaia\ ought to be able to see a near-complete orbit. If a 1 \Msun\ star is observed to have an acceleration above the limit here, but a complete orbit is not seen over \gaia's lifetime, only a compact object companion is  possible.}  
    \label{fig:BH_detectability}
    \end{center}
\end{figure}

For some systems, with \gaia\ alone, a non-degenerate companion can be ruled out by placing the limit $M_2<M_1$, since the more massive star is the brighter of the two\footnote{Exceptions to this may exist for evolved companions, but such systems are rare.}. Applying this limit to Equation \ref{eq:mass_constraint}, we find a limit on the {\it luminous} star's mass:
\begin{equation}
    M_1 \gtrsim 3 \left( \frac{ \left.V^2/R\right|_1}{10^{-2}\ {\rm cm}\ {\rm s}^{-2}} \right)^3 \left( \frac{\tau_{\rm Gaia}}{10\ {\rm years}} \right)^4 \Msun. \label{eq:mass_limit}
\end{equation}
For systems satisfying this constraint that do not complete a full orbit over \gaia's lifetime, the observed acceleration may be due to a dark companion more massive than the observed star. A luminous companion cannot simultaneously produce the observed acceleration, be less luminous than the primary, and not complete a full orbit over \gaia's lifetime. However, in such cases, a BH companion cannot be automatically assumed, as other exotic combinations may be possible (for instance the unseen companion could itself be a binary, that is faint, but with a substantial combined mass). We can only speculate that such systems will also be rare, and may themselves be astrophysically interesting (and therefore worthy of observational follow-up). Applying this limit to our population synthesis predictions, we find $\simeq$15 of the 47 BH binaries can be identified as having dark compact objects\footnote{Note that depending on the system, both neutron star and BH companions may be possible; however we consider the detection of even one such system, regardless of whether it contains a neutron star or BH, to be a significant contribution to the study of compact object binaries.}. 

Figure \ref{fig:BH_detectability} demonstrates the acceleration on a 1 \Msun\ luminous star caused by a companion object, as a function of its mass. We assume that for orbits with periods smaller than $\sim$20 years, \gaia\ will observe enough of the orbit to attempt a full fit. For longer period orbits, dark, massive companions are one possibility that can produce sufficiently large orbital accelerations (blue region).

We note that the relative orientation of an individual system ought to be irrelevant since the projected acceleration is always less than the total 3D acceleration. Therefore, any individual system identified as hosting a BH due to its high acceleration cannot be confused with a lower-mass, non-degenerate star in a circular orbit viewed at an inclined angle. 

It is also worth considering contamination by non-faint companions. Since the luminosity of a Main Sequence star scales with $M^{3.5}$, even slightly less-massive companions will contribute little to the overall luminosity of the system. However, for binaries with near-unity mass ratios, \gaia\ observes the motion of the photocenter of the system, rather than the more luminous star. The derivation of system parameters for such astrometric binaries was explored in a series of papers in the context of {\it Hipparcos} observations \citep{martin97, martin98a, martin98b}. An extension of this concept to \gaia\ is outside the scope of this work and has been addressed elsewhere \citep{penoyre20}. Here, we only comment that for unresolved stars of equal magnitude, the binary's photocenter will not display orbital motion. For stars with slightly differing masses, the photocenter moves, but the apparent acceleration is always smaller than that felt by the more massive star. Therefore our constraints defined in Equations \ref{eq:mass_constraint} and \ref{eq:mass_limit} are unaffected by the possibility of luminous companions.

Finally, we note that the limits described in Equations \ref{eq:mass_constraint} and \ref{eq:mass_limit} are derived assuming circular binaries. However, our analysis in Section \ref{sec:theory} demonstrates that an eccentric binary at periastron produces a somewhat larger acceleration than its circular equivalent. While there is a distinct possibility that faint companions could produce the acceleration observed in any individual system identified using Equations \ref{eq:mass_constraint} and \ref{eq:mass_limit} as hosting a dark, compact companion, Figure \ref{fig:eccentricity_contribution} shows that the chances of any one binary comprised of two luminous stars appearing to host a black hole are quite small; a small tail exists towards large accelerations (corresponding to the binary at periastron), but eccentric binaries typically show smaller orbital accelerations than their circular equivalent. However, binaries comprised of two luminous stars are common, while black hole binaries are extremely rare. We leave a comparative analysis determining the relative rates of the two populations to a future work.

\section{Discussion and Conclusions}
\label{sec:conclusions}

Although many previous studies have focused on the measurement of astrometric binaries by \gaia, nearly all focus on either systems in which \gaia\ observes at least one complete orbit or how \gaia\ data can be combined with other astrometric and spectroscopic data of known systems. According to \gaia's data release scenario\footnote{\url{https://www.cosmos.esa.int/web/gaia/release}}, the next data release (DR3) will contain information on astrometric binaries. Many of these will be systems with complete orbital solutions; however this data is also expected to contain a catalog of stars showing non-linear proper motions -- orbital accelerations. In this work, we seek to broaden the family of binaries relevant for \gaia, by quantifying orbital accelerations in binaries with periods longer than \gaia's lifetime. By fitting synthetic data using a Markov-Chain Monte Carlo method, we derive fitting formula that can be used to quantify \gaia's ability to measure the orbital acceleration of any particular long-period binary. Depending on its distance, we find \gaia\ can measure orbital accelerations as small as $\sim10^{-5}$ cm s$^{-2}$. 

We then calculate the barycentric orbital acceleration as a function of orbital phase for eccentric binaries in Section \ref{sec:theory}. We find that inclination effects due to the orientation of the binary's orbital plane are generally negligible. After combining with the length of time a binary spends at different orbital phases and a uniform distribution of eccentricities, we show in Figure \ref{fig:eccentricity_distribution} that orbits typically have accelerations comparable to, but somewhat smaller than, their circular equivalent. For a thermal eccentricity distribution, the orbital accelerations are typically a factor of $\simeq$3 smaller. 

For any individual system the observed acceleration is a function of the mass of the unknown companion and the orbital separation, a degeneracy which precludes a detailed fit for any individual system. However, populations of binaries ought to produce observable signals in the distribution. For example, if one knows the distribution of orbital separations of a population of long-period binaries, the distribution of companion masses can be derived from a \gaia\ sample of orbital accelerations. Or alternatively, a distribution of orbital separations can be derived if the distribution of companion masses is known. Again, eccentricity and inclination generally have only minor effects (and produce the same mean effect on all binaries anyway) so can be ignored at first order.

We then use binary population synthesis in Section \ref{sec:pop_synth} to generate a realistic population of BH binaries using our current best knowledge of binary evolution physics. After producing a sample of binaries and randomly placing them throughout the Milky Way according to a distribution and rate consistent with detailed evolution models of our Galaxy, we find that \gaia\ ought to observe $\simeq$47 accelerating stars with BH companions. However, most of these systems are unlikely to be uniquely identified by \gaia. The degeneracy between mass and orbital separation means the same acceleration produced by any individual BH could also be produced by a faint, low-mass star with a smaller orbital separation. Nevertheless, for sufficiently small separations, \gaia\ will observe a complete orbit during its lifetime; if \gaia\ only sees an incomplete arc for a particular star, a lower limit on the companion mass can be derived. Using our binary population synthesis, we estimate that 15 BH binaries have sufficiently high lower limits on the companion masses that they can be unambiguously identified as hosting compact object companions.

These estimates are produced under the assumption of one particular parameterization of binary evolution. In \citet{Breivik2018}, we demonstrate that the population of astrometric binaries containing a BH depends strongly on which parameters we choose to adopt; in that work, we focused on two separate BH mass functions from \citet{fryer2012}. Although we do not explicitly compare the two different prescriptions here, each will certainly affect the resulting yield of detached BH systems that \gaia\ will detect. A clear example for the case at present comes from how we model the kicks that BHs receive at birth. We adopt a model in which BH natal kicks are reduced due to fallback. For sufficiently massive BHs, these natal kicks become irrelevant; a 10 \Msun\ BH with an orbital separation of 10 au around a luminous star has an orbital velocity of $\simeq$30 km s$^{-1}$. However, if we adopt a model in which BHs receive a somewhat larger kick, similar in magnitude or larger at birth, a fraction of these will disrupt, reducing \gaia's yield. Although a detailed comparison to different population synthesis prescriptions is outside the scope of this work, the case of BH kicks serves as an example which demonstrates, at least in principle, how a population of wide-period BH binaries can be used to constrain uncertain binary evolution physics.

Since the latest data release from \gaia\ (DR2) does not contain information about stellar binaries, the current best sample of stars with measured accelerations is that from \citet{Brandt2018}, who compares the proper motions from \gaia\ to that from \hipparcos. With its 24-year baseline, for stars detected by \hipparcos\ this catalog forms an excellent resource. Our scaling with time baseline in Equation \ref{eq:detectability_sigma} shows that sources detected by \gaia\ alone, with its 10-year lifetime, will have a factor of $\simeq$6 degradation in acceleration measurement. However, this is more than made up for by the factor of $\sim10^2$ improvement in \gaia's astrometric precision over its predecessor; our simple estimate suggests that \gaia\ will be able to measure astrometric accelerations a factor of $\sim$20 better than has been possible previously, and for $\sim 10^9$ stars. Therefore, the 10$^5$ stars in the catalog from \citet{Brandt2018} represent the tip of a much larger iceberg soon to be available.

\acknowledgements

We thank the referee for a careful reading of the manuscript and for constructive comments which greatly improved its quality. J.J.A.\ acknowledges support from the Danish National Research Foundation (DNRF132) and from CIERA and Northwestern University through a Postdoctoral Fellowship. K.B.\ is grateful for support from the Jeffrey L. Bishop Fellowship. This work was initiated and performed in part at the Aspen Center for Physics, which is supported by National Science Foundation grant PHY-1607611. The simulations within this work were run on the Metropolis HPC Facility at the CCQCN Center of the University of Crete, supported by the European Union Seventh Framework Programme (FP7-REGPOT-2012-2013-1) under grant agreement no.\ 316165. SC acknowledges support from the Department of Atomic Energy, Government of India, under project no. 12-R\&D-TFR-5.02-0200.

\software{{\tt astropy} \citep{astropy}, {\tt SKYCALC} (J.\ Thorstensen, private communication), {\tt emcee} \citep{Foreman-Mackey2013}, {\tt NumPy} \citep{numpy}, {\tt SciPy} \citep{scipy}, {\tt matplotlib} \citep{matplotlib}}

\bibliographystyle{aasjournal}
\bibliography{gaia}

\end{document}